\documentclass[aps,twocolumn,nofootinbib,superscriptaddress,preprintnumbers,pra,10pt]{revtex4-1}
\usepackage{float}
\usepackage{colortbl}
\usepackage{pifont}
\newcommand{\cmark}{\text{\ding{51}}}
\newcommand{\xmark}{\text{\ding{55}}}
\usepackage{amsmath,booktabs,amssymb
}
\usepackage{graphicx,color}
\usepackage[usenames,dvipsnames]{xcolor}
\usepackage{dsfont} 
\usepackage[colorlinks,bookmarksopen,bookmarksnumbered,pdfstartview=FitH,citecolor=SBred, linkcolor=SBred,urlcolor=SBred]{hyperref}
\usepackage{graphicx}
\usepackage{ulem} 
\usepackage{enumitem}

\newcommand\Tstrut{\rule{0pt}{2.6ex}}       
\newcommand\Bstrut{\rule[-0.9ex]{0pt}{0pt}} 
\newcommand{\TBstrut}{\Tstrut\Bstrut} 

\definecolor{Gray}{gray}{0.95}
\definecolor{RGray}{gray}{0.85}
\definecolor{CGray}{gray}{0.92}

\definecolor{color1}{rgb}{0.9,.4,.2}
\definecolor{color2}{rgb}{0.3,.6,.7}
\definecolor{color3}{rgb}{0.7,.2,.7}

\newcommand{\mew}{\mu_{\text{\tiny{EW}}}}

\definecolor{SBred}{rgb}{0.6471, 0.1098, 0.1882}

\def\a23{\alpha_{23}}

\newcommand{\be}{\begin{equation}}
\newcommand{\ee}{\end{equation}}

\newcommand{\dsix}{{\tt DsixTools}}

\newcommand{\C}{{\cal C}}

\newcommand{\Op}[1]{{\mathcal O}_{#1}}

\def\beqn#1{\begin{equation}\label{#1}}
\def\eeqn{\end{equation}}
\def\beqa#1{\begin{eqnarray}\label{#1}}
\def\eeqa{\end{eqnarray}}

\begin{document}

\preprint{LMU-ASC 25/17}
\preprint{IFIC/17-20}

\title{Gauge-invariant implications of the LHCb measurements\\[2mm]
on Lepton-Flavour Non-Universality}
 
\author{Alejandro Celis}
 \affiliation{Ludwig-Maximilians-Universit\"at M\"unchen, 
   Fakult\"at f\"ur Physik,\\
   Arnold Sommerfeld Center for Theoretical Physics, 
   80333 M\"unchen, Germany}
\author{Javier Fuentes-Mart\'{\i}n}
\affiliation{Instituto de F\'{\i}sica Corpuscular, Universitat de Val\`encia - CSIC, E-46071 Val\`encia, Spain}
\author{Avelino Vicente}
\affiliation{Instituto de F\'{\i}sica Corpuscular, Universitat de Val\`encia - CSIC, E-46071 Val\`encia, Spain}
\author{Javier Virto}
\affiliation{\mbox{Albert Einstein Center for Fundamental Physics, Institute for Theoretical Physics,}\\
University of Bern, CH-3012 Bern, Switzerland}


\begin{abstract}
\vspace{5mm}

We study the implications of the recent measurements of $R_K$ and
$R_{K^\ast}$ by the LHCb collaboration. We do that by adopting a
model-independent approach based on the Standard Model Effective Field
Theory (SMEFT), in which the dominant new physics effects are encoded
in the coefficients of dimension-six operators respecting the full Standard Model (SM) gauge symmetry. After providing
simplified expressions for $R_K$ and $R_{K^\ast}$, we determine the
implications of the recent LHCb results for these observables on the
coefficients of the SMEFT operators at low and high energies. We also 
take into account all $b\to s \ell\ell$ data, which combined lead to effective New Physics (NP) scenarios with SM pulls in excess of 5~$\sigma$. Thus the operators discussed in this paper would be the first dimension-six terms in the SM Lagrangian to be detected experimentally.   Indirect constraints on these operators are also discussed. The results of this paper transcend the singularity of the present situation, and set a standard for future analyses in $b\to s$ transitions when the NP is assumed to lie above the electroweak scale.

\vspace{3mm}
\end{abstract}

\maketitle

\allowdisplaybreaks

\section{Introduction}\label{sec:intro}

An absolute priority in particle physics is to detect and to measure the
effects of dimension-six terms in the SM effective Lagrangian, which according to our current understanding {\it must} be there if the SM is only valid up to a
physical cut-off scale $\Lambda > \mew$, where $\mew$
is the scale of electroweak physics. These effects will be suppressed by a factor $\mew^2/\Lambda^2$, so if $\Lambda$ is very large, we need to choose wisely where to look.

In this sense, one of the most important features of the SM is
lepton-flavour universality (LFU), the interactions between gauge
bosons and leptons being exactly the same for different lepton
families. This central prediction can be easily modified by dimension-six terms, and can be tested precisely by measuring observables such
as the $R_{K^{(\ast)}}$ ratios, defined as~\cite{Hiller:2003js}
\begin{align}
R_{K^{(\ast)}} = \frac{ \Gamma(B \rightarrow K^{(\ast)} \mu^+ \mu^-)}{\Gamma(B \rightarrow K^{(\ast)} e^+ e^-)} \, ,
\end{align}
measured in specific dilepton invariant mass squared ranges $q^2 \in
[q^2_{\rm min}, q^2_{\rm max}]$. In the absence of large LFU-violating NP, hadronic uncertainties cancel to very good approximation in these ratios,
which become robust tests of the SM. Up to lepton-mass effects, these ratios should be
very approximately equal to one in the absence of LFU-violating new physics.

In 2014 the LHCb collaboration reported a measurement of the ratio $R_K$ in the region $[1,6]$~GeV$^2$~\cite{Aaij:2014ora}, finding a value significantly lower than one. Very recently, a similar measurement of the
ratio $R_{K^\ast}$~\cite{LHCbtalk} in two $q^2$ bins indicates a similar violation of LFU: 
\begin{align}
R_K &= 0.745^{+0.090}_{-0.074}\pm0.036    \,, \quad
q^2 \in [1,6]~\text{GeV}^2 \,, \nonumber \\[0.2cm]
R_{K^\ast} &= 0.660^{+0.110}_{-0.070}\pm0.024    \,, \quad
q^2 \in [0.045,1.1]~\text{GeV}^2\,, \nonumber \\[0.2cm]
R_{K^\ast} &= 0.685^{+0.113}_{-0.069}\pm0.047    \,, \quad
q^2 \in [1.1,6.0]~\text{GeV}^2 \,. 
\end{align}
When these experimental results are compared to their SM predictions~\cite{Descotes-Genon:2015uva},
\begin{align}
R_K^{\rm{SM}} &= 1.00 \pm 0.01   \,, \quad
q^2 \in [1,6]~\text{GeV}^2 \,, \nonumber \\[0.2cm]
R_{K^\ast}^{\rm{SM}} &= 0.92 \pm 0.02    \,, \quad
q^2 \in [0.045,1.1]~\text{GeV}^2 \,,\nonumber \\[0.2cm]
R_{K^\ast}^{\rm{SM}} &= 1.00 \pm 0.01    \,, \quad
q^2 \in [1.1,6.0]~\text{GeV}^2 \,,
\end{align}
one concludes that the LHCb measurements represent deviations from the
SM at the $2.6\,\sigma$ level in the case of $R_K$, $2.2\,\sigma$ for
$R_{K^\ast}$ in the low-$q^2$ region, and $2.4\,\sigma$ for
$R_{K^\ast}$ in the central-$q^2$ region.\footnote{An important issue in these observables is related to electromagnetic effects. Given the experimental treatment,
the residual theory error
has been estimated to be of $\mathcal{O}(1\%)$ in Refs.~\cite{Guevara:2015pza,Bordone:2016gaq}.} Recent studies analysing these new measurements  in terms of NP models and the Weak Effective Theory (WET) can be found in Refs.~\cite{Capdevila:2017bsm,Altmannshofer:2017yso,DAmico:2017mtc,Hiller:2017bzc,Geng:2017svp,Ciuchini:2017mik}.

Also recently, the Belle collaboration found slight differences between the electron
and muon channels in their lepton-flavour-dependent (but isospin averaged) angular analysis of $B\to K^\ast \ell^+\ell^-$ \cite{Wehle:2016yoi}, most notably in the pioneering measurement of the clean observables $Q_4$ and $Q_5$~\cite{Capdevila:2016ivx}. Although the individual statistical significance of these discrepancies is not sufficient to
claim the discovery of LFU violation (LFUV), their combination
constitutes an intriguing set of anomalies.

New physics causing LFUV in these ratios would be expected to manifest itself also in $b \to s \ell^+ \ell^-$ ($\ell = \mu$ or $e$) decay observables such as branching ratios and angular distributions.  Interestingly, current data on $b \to s \mu^+ \mu^-$ transitions show departures with respect to the SM predictions too~\cite{Aaij:2013qta,Descotes-Genon:2013wba,Aaij:2015oid}, which are consistent with the anomaly in $R_K$~\cite{Alonso:2014csa,Hiller:2014ula,Ghosh:2014awa}.
Global analyses of $b \to s \mu^+ \mu^-$ data within the WET hint to new physics scenarios that can accommodate also the observed LFUV in $R_K$ and $R_{K^\ast}$~\cite{Hurth:2014vma,Altmannshofer:2014rta,Descotes-Genon:2015uva}.  

Many models have been proposed to address
the $b\to s$ anomalies (including $R_K$).
These models involve a $Z^{\prime}$ boson from an extended gauge
group~\cite{Buras:2013dea,Buras:2013qja,Altmannshofer:2014cfa,Crivellin:2015lwa,Crivellin:2015mga,Sierra:2015fma,Celis:2015ara,Belanger:2015nma,Celis:2015eqs,Falkowski:2015zwa,Allanach:2015gkd,Alonso:2015sja,Bauer:2015knc,Fajfer:2015ycq,Barbieri:2015yvd,Hati:2016thk,Deppisch:2016qqd,Das:2016vkr,Chiang:2016qov,Kim:2016bdu,Boucenna:2016wpr,Boucenna:2016qad,Celis:2016ayl,Altmannshofer:2016jzy,Crivellin:2016ejn,C.:2017yqh,Bnaerjee:2017zph,Ko:2017quv,Bhatia:2017tgo,Cline:2017lvv,Datta:2017pfz,Ko:2017yrd,Ko:2017lzd,Chen:2017hir,Alok:2017jgr,Ahmed:2017vsr},
leptoquarks (or R-parity violating
supersymmetry)~\cite{Hiller:2014yaa,Biswas:2014gga,Gripaios:2014tna,Varzielas:2015iva,Becirevic:2015asa,Sahoo:2015wya,Sahoo:2015qha,Sahoo:2015fla,Pas:2015hca,Huang:2015vpt,Chen:2016dip,Deshpand:2016cpw,Becirevic:2016oho,Becirevic:2016yqi,Sahoo:2016pet,Hiller:2016kry,Bhattacharya:2016mcc,Duraisamy:2016gsd,Cheung:2016frv,Popov:2016fzr,Barbieri:2016las,Cox:2016epl,Crivellin:2017zlb,Alok:2017jgr},
a massive resonance from a strong
dynamics~\cite{Niehoff:2015bfa,Niehoff:2015iaa,Carmona:2015ena,Greljo:2015mma,Buttazzo:2016kid}
or Kaluza-Klein
excitations~\cite{Megias:2016bde,GarciaGarcia:2016nvr,Megias:2017dzd,Megias:2017ove}. Refs.~\cite{Gripaios:2015gra,Arnan:2016cpy,Hu:2016gpe,Arnan:2017lxi}
have explored renormalizable models that explain $R_K$ at the one-loop
level, while the MSSM with R-parity conservation was considered in
Ref.~\cite{Mahmoudi:2014mja}. 

In this work we interpret the new LHCb indications of LFUV in a
model-independent way using the SMEFT
~\cite{Buchmuller:1985jz,Grzadkowski:2010es}.  This framework
provides the most general description once we assume that the SM is
valid at low energies and the NP decouples at a scale much higher than
the EW scale.  The interpretation in terms of the SMEFT
allows for a more transparent connection to possible ultraviolet (UV)
scenarios as it incorporates the full electroweak gauge symmetry (see
for instance~\cite{Alonso:2014csa}). We point out that the difference
with respect to the analyses of
Refs.~\cite{Bhattacharya:2014wla,Calibbi:2015kma,Feruglio:2016gvd,Bordone:2017anc} is
that here we do not assume that only operators with third generation
fermions are generated, or any underlying flavour symmetry.

We will start by
providing simplified analytical expressions for the observables of
interest, as well as for the SMEFT Wilson coefficients (WCs) at low
and high energies. These expressions can be of great value to guide
the model building efforts.  With these expressions at hand, we
determine the implications of the LHCb measurements, not only on the
coefficients of the SMEFT operators at low energies, but also on their
values at the high-energy scale where they are generated by
the decoupling of some unknown heavy degrees of freedom. For this
purpose we will make use of \dsix~\cite{Celis:2017hod}, a {\tt Mathematica} package for the
handling of the dimension-six SMEFT~\cite{Celis:2017hod}, which implements the
complete one-loop Renormalization Group Equations (RGEs) of the SMEFT.
This package will also allow us to consider the generation of other
(unwanted) effective operators at low energies due to the RGE
evolution of the SMEFT operators and find which of these imply
relevant constraints on the scenarios that explain the LHCb measurements.

The rest of this letter is organized as follows: in Sec.~\ref{sec:ops}
we introduce the relevant SMEFT and WET operators. In
Sec.~\ref{sec:formulas} we find simple analytical expressions for the
$R_K$ and $R_{K^\ast}$ ratios. In Sec.~\ref{sec:implications} we
analyse the implications of the LHCb measurements and identify the
SMEFT scenarios that can accommodate them.  
RGE effects from the high-energy scale of the new dynamics to the electroweak scale are discussed in Sec.~\ref{sec:constraints}. Finally, we
conclude and discuss further implications in
Sec. \ref{sec:summary}.

\section{Effective Field Theory}
\label{sec:ops}

At energies relevant for the $B$-meson decays, NP effects can be described generically in terms of the Weak Effective Theory (WET).  Semileptonic $b \to s$ transitions involve the effective weak Hamiltonian
\begin{equation} \label{eq:WeakHeff}
  \mathcal{H}_{\rm{eff}} \supset- \frac{  4 G_F }{\sqrt{2}}   \frac{\alpha}{4 \pi}   \lambda_{t}^{sb} \sum_{i} \C_{i}   \Op{i} \,,
\end{equation}
where $\lambda_{t}^{ij} = V_{ti}^* V_{tj}$, with $V$ the
  Cabibbo-Kobayashi-Maskawa (CKM) matrix, and $\lambda_{t}^{sb} \sim
-0.04$~\cite{Charles:2004jd}.  The most relevant operators for the present purpose are
the semileptonic operators
\begin{align}
\begin{aligned}
\Op{9} &= (  \bar s \gamma_{\alpha} P_L b  ) (  \bar \ell \gamma^{\alpha} \ell )  \,, \qquad \Op{9}^{\prime} = (  \bar s \gamma_{\alpha}   P_R b  )( \bar \ell \gamma^{\alpha} \ell ) \,, \\
\Op{10} &= (  \bar s \gamma_{\alpha} P_L b  ) (  \bar \ell \gamma^{\alpha}  \gamma_5 \ell )  \,, \quad \Op{10}^{\prime} = (  \bar s \gamma_{\alpha}   P_R b  )( \bar \ell \gamma^{\alpha}  \gamma_5 \ell ) \,, \nonumber
\end{aligned}
\end{align}
and the dipole operator
\begin{equation}
\Op{7} = \frac{m_b}{e} (  \bar s \sigma_{\alpha \beta}   P_R b  ) \, F^{\alpha \beta} \, , \nonumber
\end{equation}
with $m_b$ the $b$-quark mass and $F^{\alpha \beta}$ the electromagnetic field-strength tensor. 

Assuming that the SM degrees of freedom are the only ones present below a certain mass scale $\Lambda \gg M_W$ where NP decouples, one can describe deviations from the SM in a general way using the SMEFT.  Dominant NP effects in $b \to s$ transitions are expected to be parametrized by effective operators of canonical dimension six
\begin{equation} \label{eq:SMEFT}
\mathcal{L}_{\rm SMEFT} \supset \frac{1}{\Lambda^2} \sum_k\C_k Q_k \, .
\end{equation}
Here $\C_k$ are the Wilson coefficients of the dimension-six $Q_k$
operators. In this letter we will adopt the so-called \textit{Warsaw basis} for the dimension-six operators
\cite{Grzadkowski:2010es}.

\begin{table}
\centering
\renewcommand*{\arraystretch}{2.0}
\vspace{10pt}
\begin{tabular}{@{}cccl@{}}
\hline
SMEFT operator & Definition & Matching & Order   \Bstrut\\
\hline
$[Q_{\ell q}^{(1)}]_{aa23}$  & $\left( \bar \ell_a \gamma_\mu \ell_a \right) \left( \bar q_2 \gamma^\mu q_3 \right)$   & $\Op{9,10}$ & Tree\TBstrut\\
$[Q_{\ell q}^{(3)}]_{aa23}$  & $\left( \bar \ell_a \gamma_\mu \tau^I \ell_a \right) \left( \bar q_2 \gamma^\mu \tau^I q_3 \right)$  & $\Op{9,10}$ & Tree\TBstrut\\
$[Q_{qe}]_{23aa}$          & $\left( \bar q_2 \gamma_\mu q_3 \right) \left( \bar e_a \gamma^\mu e_a \right)$  &  $\Op{9,10}$ & Tree \TBstrut\\
$[Q_{\ell d}]_{aa23}$       & $\left( \bar \ell_a \gamma_\mu \ell_a \right) \left( \bar d_2 \gamma^\mu d_3 \right)$   & $\Op{9,10}^\prime$ & Tree\TBstrut\\
$[Q_{e d}]_{aa23}$         & $\left( \bar e_a \gamma_\mu e_a \right) \left( \bar d_2 \gamma^\mu d_3 \right)$    & $\Op{9,10}^\prime$ & Tree\Tstrut\\
$[Q_{\varphi \ell}^{(1)}]_{aa\phantom{d}}$         & $\left(\varphi^\dagger i\overleftrightarrow D_\mu\varphi\right)\left(\bar \ell_a \gamma^\mu \ell_a \right)$    & $\Op{9,10}$ & 1-loop\Tstrut\\
$[Q_{\varphi \ell}^{(3)}]_{aa\phantom{d}}$         & $\left(\varphi^\dagger i\overleftrightarrow D_\mu^I\varphi\right)\left(\bar \ell_a \gamma^\mu\tau^I \ell_a \right)$    & $\Op{9,10}$ & 1-loop\Tstrut\\
$[Q_{\ell u}]_{aa33}$         & $\left( \bar \ell_a \gamma_\mu \ell_a \right) \left( \bar u_3 \gamma^\mu u_3 \right)$    & $\Op{9,10}$ & 1-loop\Tstrut\\
$[Q_{\varphi e}]_{aa\phantom{d}}$         & $\left(\varphi^\dagger i\overleftrightarrow D_\mu\varphi\right)\left(\bar e_a \gamma^\mu e_a \right)$    & $\Op{9,10}$ & 1-loop\Tstrut\\
$[Q_{e u}]_{aa33}$         & $\left( \bar e_a \gamma_\mu e_a \right) \left( \bar u_3 \gamma^\mu u_3 \right)$    & $\Op{9,10}$ & 1-loop\Tstrut\\
\hline
\end{tabular}
\caption{List of relevant operators (see Ref.~\cite{Grzadkowski:2010es} for definitions) that contribute to the matching to $\C^{(\prime)}_{9,10}$, either at tree-level or through one-loop running. The index $a = \mu, e$ denotes the flavour of the lepton.
\label{tab:matching}}
\end{table}

One can match the SMEFT operators onto the operators in~Eq.~\eqref{eq:WeakHeff}. The relevant matching conditions at the EW
scale $\mew \sim  \mathcal{O}(M_W)$ are given
by~\cite{Alonso:2014csa,Aebischer:2015fzz} (with $a = e,\mu$):
\begin{align}  \label{eqmatchgre}
\C_{9a}^{\rm NP} &= \frac{\pi}{\alpha \lambda_{t}^{sb}} \frac{v^2}{\Lambda^2}
\bigg\{
\big[\tilde\C_{\ell q}^{(1)}\big]_{aa23}
+ \big[\tilde\C_{\ell q}^{(3)}\big]_{aa23}
+ \big[\tilde\C_{qe}\big]_{23aa} \bigg\} \,,  \nonumber  \\[3mm]
\C_{10a}^{\rm NP} &= - \frac{\pi}{\alpha \lambda_{t}^{sb}} \frac{v^2}{\Lambda^2}
\bigg\{
\big[\tilde\C_{\ell q}^{(1)}\big]_{aa23}
+ \big[\tilde\C_{\ell q}^{(3)}\big]_{aa23}
- \big[\tilde\C_{qe}\big]_{23aa} \bigg\} \,, \nonumber \\[3mm]
\C_{9a}^{\prime} &= \frac{\pi}{\alpha \lambda_{t}^{sb}} \frac{v^2}{\Lambda^2}
\bigg\{
\big[\tilde\C_{\ell d}\big]_{aa23}
+ \big[\tilde\C_{ed}\big]_{aa23} \bigg\}\,, \nonumber \\[3mm]
\C_{10a}^{\prime} &= - \frac{\pi}{\alpha \lambda_{t}^{sb}} \frac{v^2}{\Lambda^2}
\bigg\{
\big[\tilde\C_{\ell d}\big]_{aa23}
- \big[\tilde\C_{ed}\big]_{aa23} \bigg\}\ .
\end{align}
We point out that in these expressions we have only included operators that can give rise to LFUV. These matching conditions are also summarized in
Table~\ref{tab:matching}, where the operators of the SMEFT are defined.  We also show in this table the operators that contribute via one-loop running, 
but leave out a few others that contribute finite terms to the matching.  Here we implicitly assume that the Wilson coefficients $\C_{9\mu}^{\rm
  NP}$, $\tilde\C_{\ell q}^{(1)}$, etc., are defined at the matching
scale $\mew$, {\it i.e.} $\C_{9\mu}^{\rm NP}\equiv \C_{9\mu}^{\rm
  NP}(\mew)$, etc. The tilde over the SMEFT Wilson coefficients
denotes that they are given in the fermion mass basis (see Appendix~\ref{secfba}).

The dipole operator $\Op{7}$ receives tree level matching contributions from $[Q_{dB}]_{23}  =  (   \bar q_{2}   \sigma^{\mu \nu}   d_3  )  \varphi   B_{\mu \nu}$  and 
$[Q_{dW}]_{23}  =  (   \bar q_{2}   \sigma^{\mu \nu}   d_3  )  \tau^{I} \varphi   W_{\mu \nu}^{I}$, both dipole operators of the SMEFT~\cite{Alonso:2014csa,Aebischer:2015fzz}.   Assuming that the underlying UV model is a weakly coupled
gauge theory, all the operators in Table~\ref{tab:matching} contributing to $\Op{9,10}^{(\prime)}$ are potentially
generated at tree-level by the new physics~\cite{Arzt:1994gp}. In contrast, the SMEFT dipole
operators contributing to $\Op{7}$ would be loop-generated~\cite{Arzt:1994gp}.

\section{Formulas for $\mathbf{R_K}$ and $\mathbf{R_{K^\ast}}$}
\label{sec:formulas}

For the phenomenological discussion we derive approximate formulas for $R_K$ and $R_{K^\ast}$ in terms of the relevant WCs.  These formulas are obtained with the same
approach as Ref.~\cite{Descotes-Genon:2015uva}, but neglecting terms that are not important for the present discussion, and linearising in the NP coefficients.
We find:
\begin{widetext}   
\begin{align}
[R_{K}]_{[1,6]}  \simeq&\; 1.00(1)  + 0.230 (\C_{9 \mu-e}^{\rm{NP}} +\C_{9 \mu-e}^{\prime}   ) - 0.233(2) (\C_{10 \mu-e}^{\rm{NP} }  +\C_{10 \mu-e}^{\prime} ) \,,  \nonumber \\[0.3cm]
[R_{K^\ast}]_{[0.045, 1.1]} \simeq&\; 0.92(2)  + 0.07(2)\C_{9 \mu-e}^{\rm{NP}}  - 0.10(2)\C_{9 \mu-e}^{\prime}     - 0.11(2)\C_{10 \mu-e}^{\rm{NP}}    + 0.11(2)\C_{10 \mu-e}^{\prime} + 0.18(1)\C_{7}^{\rm{NP} }     \,,   \nonumber \\[0.3cm]
[R_{K^\ast}]_{[1.1,6]}   \simeq&\; 1.00(1) + 0.20(1)\C_{9 \mu-e}^{\rm{NP}}   - 0.19(1) \C_{9 \mu-e}^{\prime}     - 0.27(1) \C_{10 \mu-e}^{\rm{NP}}     + 0.21(1)\C_{10 \mu-e}^{\prime}  \,.
\label{eq:obs}
\end{align}
\bigskip
\end{widetext}
All WCs in \eqref{eq:obs} are assumed to be defined at the $\mu_b \sim 4.8$~GeV scale.   The notation $\C_{9 \mu-e}^{\rm{NP}} \equiv \C_{9 \mu}^{\rm{NP}} - \C_{9 e}^{\rm{NP}}$ (and similarly for the other WCs) has been used.   We have
linearised the dependence with respect to the WCs in these formulas,
assuming that contributions from dimension-eight SMEFT operators
interfering with the SM as well as contributions from dimension-six
SMEFT operators squared are both negligible.\footnote{Some of our results will be presented using the exact formulas, without linearising in the NP coefficients.  In these cases we are neglecting the effect of dimension-eight SMEFT operators
interfering with the SM. } 

Semileptonic four-fermion operators enter in these formulas in combinations $\mu-e$ to very good accuracy, so that they will source LFUV effects only when the values of these coefficients differ for muons and electrons.     The dipole operator $\Op{7}$ is relevant in the low bin of $R_{K^\ast}$, where it enters due to lepton mass effects.   For $[R_{K}]_{[1,6]}$ and $[R_{K^\ast}]_{[1.1,6]}$ these formulas are in reasonably good agreement with those in \cite{Hiller:2014ula}.  The scalar operators in the WET $\Op{S}^{(\prime)} = m_b  (  \bar s P_{R,L} b )(\bar \ell \ell)$ do not enter in these formulas since they do not interfere with the SM.         Contributions from the pseudo-scalar operators $\Op{P}^{(\prime)} = m_b  (  \bar s P_{R,L} b )(\bar \ell \gamma_5  \ell)$ and the chirality-flipped dipole operator $\Op{7}^{\prime} = m_b/e (  \bar s \sigma_{\alpha \beta}   P_L b  ) \, F^{\alpha \beta} $ to these ratios are found to be very suppressed and are therefore not included in \eqref{eq:obs}.

\begin{figure}
\centering
\includegraphics[width=8.7cm]{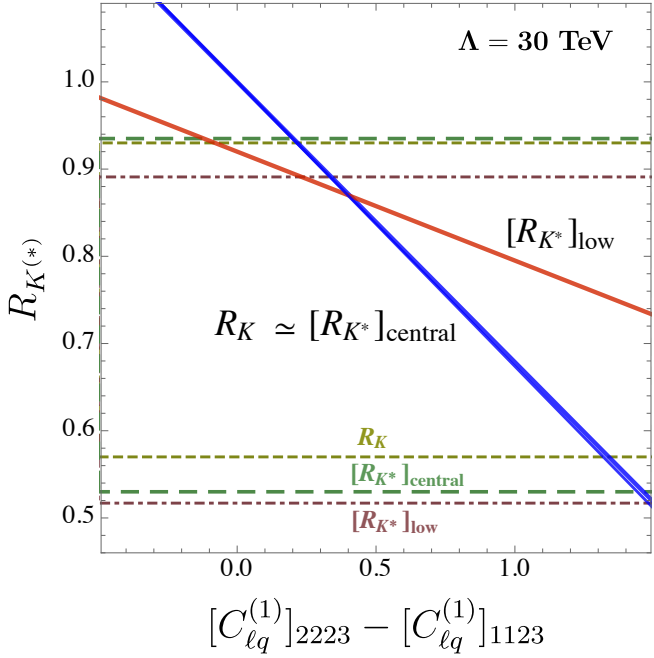} 
\caption{\small \sf  Predictions for $R_K$ and $R_{K^*}$ as a function of the SMEFT Wilson coefficient $\C_{\ell q}^{(1)}$ with $\Lambda = 30$~TeV.    The experimental ranges for $R_K$ and $R_{K^*}$ at $95\%$~CL are also shown for comparison.   }
\label{fig:bounds2}
\end{figure}

\section{Explaining LHCb measurements}
\label{sec:implications}

We investigate the implications of the LHCb measurements by considering the measured $95\%$~confidence level (CL) intervals $[R_{K^\ast}]_{[0.045, 1.1]} \in [0.517,0.891]$ and $[R_{K^\ast}]_{[1.1,6]} \in [0.530,0.935]$~\cite{LHCbtalk}. For $R_K$ we take the experimental measurement in \cite{Aaij:2014ora} and symmetrize the error, adding statistical and systematic errors in quadrature, obtaining $R_K \in [0.57,0.93]$ at $95\%$~CL.  

We consider only one or two of the relevant SMEFT operators at a time, and analyse which of these scenarios are able to accommodate the measurements of $R_K$ and $R_{K^*}$.   The SMEFT WCs are now assumed to be defined at the EW scale.

We start with {\textit{single-operator scenarios}}.  Note that the effect of the dipole operator $\Op{7}$ on the low-$q^2$ bin of $R_{K^*}$ is very small given the bound it receives from $b \to s \gamma$ transitions ($-0.05  \lesssim \C_{7}^{\rm{NP} } \lesssim 0.08 $ at $3\sigma$~\cite{Descotes-Genon:2015uva}).   The deviations from the SM in these three observables must then be caused mainly by the four-fermion semileptonic operators of the WET.    The only possibility to accommodate the data with a single operator is:\\[4mm]
\noindent $\blacktriangleright$ $\C_{\ell q}^{(1,3)} \rightarrow  \,\C_{9 \mu-e}^{\rm{NP}}  = -\C_{10 \mu-e}^{\rm{NP}} $ \,: these scenarios accommodate the experimental measurements of $R_K$ and $R_{K^*}$ for $\C_{9 \mu-e}^{\rm{NP}}  \lesssim - 0.2$, corresponding to $\C_{\ell q}^{(1,3)} \gtrsim 0.3$ with $\Lambda= 30$~TeV, see Figure~\ref{fig:bounds2}.      \\[3mm] 
All the other operators fail:\\[2mm]
\noindent $\blacktriangleright$ $\C_{\ell d}$  $\rightarrow \, \C_{9 \mu-e}^{\prime}  = -\C_{10 \mu-e}^{\prime}$\,: gives rise to $R_{K^\ast} >1 $ in the central-bin when $R_K < 1$.  $R_{K^\ast}$ in the low-bin is also above the experimental range when $R_K < 1$.\\[3mm]
\noindent $\blacktriangleright$  $\C_{ed}$   $\rightarrow \, \C_{9 \mu-e}^{\prime}  = \C_{10 \mu-e}^{\prime}$\,: has a very small effect on $R_K$. For reasonable values of the WC it holds $R_K \simeq R_K^{\rm SM}$. Furthermore when $R_{K^\ast} < 1$ in both bins, $R_K > 1$.\\[3mm]
\noindent $\blacktriangleright$ $\C_{qe}$  $\rightarrow \, \C_{9 \mu-e}^{\rm{NP}} =\C_{10 \mu-e}^{\rm{NP}}$\,: has a very small effect on $R_K$. For reasonable values of the WC it holds $R_K \simeq R_K^{\rm SM}$.\\[2mm]

We now consider {\textit{two-operator scenarios}}.  In this case it is possible to accommodate the hints of LFUV in $R_K$ and $R_{K^*}$ with:\\

\noindent $\blacktriangleright$  $\C_{\ell q}^{(1,3)}  \,, \,\C_{qe} \rightarrow \, \C_{9 \mu-e}^{\rm{NP}}  \,, \,\C_{10 \mu-e}^{\rm{NP}}$  \\[0.2cm]
\noindent $\blacktriangleright$  $\C_{\ell q}^{(1,3)}  \,, \,\C_{\ell d} \rightarrow \,  \C_{9 \mu-e}^{\rm{NP}} = -\C_{10 \mu-e}^{\rm{NP}}   \,,  \C_{9 \mu-e}^{\prime}  = -\C_{10 \mu-e}^{\prime}$ \\[0.2cm]
\noindent $\blacktriangleright$  $\C_{\ell q}^{(1,3)}   \,, \,\C_{ed} \rightarrow \,   \C_{9 \mu-e}^{\rm{NP}}  = -\C_{10 \mu-e}^{\rm{NP}} $  \,,  $\C_{9 \mu-e}^{\prime}  = \C_{10 \mu-e}^{\prime}$  \\[0.2cm]
\noindent $\blacktriangleright$  $\C_{\ell q}^{(1)}  \,, \,\C_{\ell q}^{(3)} \rightarrow  \,\C_{9 \mu-e}^{\rm{NP}}  = -\C_{10 \mu-e}^{\rm{NP}} $\\[0.1cm]
%

\begin{figure}
\centering
\includegraphics[width=8.6cm]{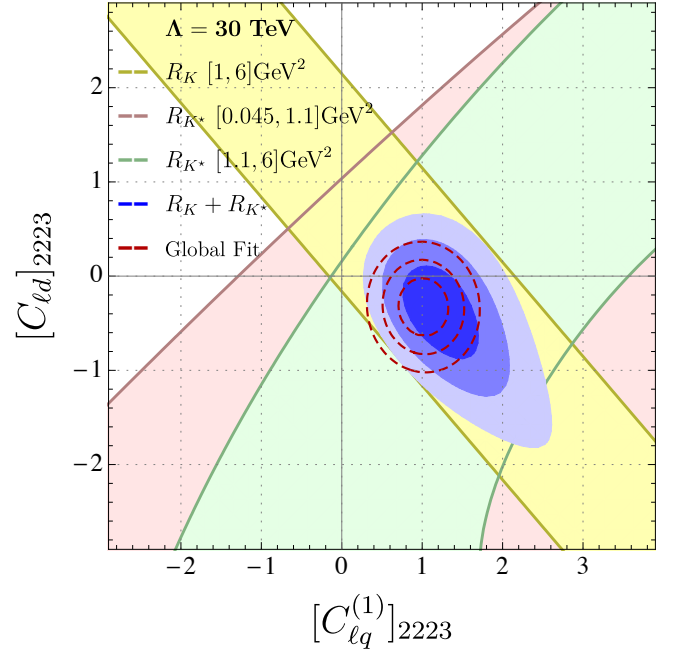} 
\caption{\small \sf  Constraints on the SMEFT Wilson coefficients $\C_{\ell q}^{(1)}$ and $\C_{\ell d}$ with $\Lambda = 30$~TeV, assuming no NP in the electron modes.  The individual constraints from $R_K$ and $R_{K^*}$ at the $3 \sigma$ level are represented by filled bands.   The combined fit to $R_K$ and $R_{K^*}$ is shown in blue (1,2 and 3 $\sigma$ contours).  The result of a global fit with all $b\to s\ell^+\ell^-$ data included in~\cite{Capdevila:2017bsm} is shown in a similar way as red dashed contours.  }
\label{fig:bounds}
\end{figure}

The bounds obtained for the WCs in the scenario of $\C_{\ell q}^{(1)}$ and $\C_{\ell d}$ are shown in Figure~\ref{fig:bounds} for illustration, assuming no NP in the electron modes.\footnote{Here we have used the exact expressions for the observables, without linearising in the NP coefficients.} The situation would be the same if we replace $\C_{\ell q}^{(1)}$ by $\C_{\ell q}^{(3)}$.    In order to accommodate the anomalies one needs a positive NP contribution to $\C_{\ell q}^{(1)}$.  The bound obtained on $\C_{\ell d}$ arises because the measurements are compatible with $[R_{K^*}]_{\rm{central}}/R_K \simeq 1$ and this double ratio is mainly sensitive to $\C_{\ell d}$.\footnote{See~\cite{Hiller:2014ula} for a discussion of this double ratio and similar observables within the WET.}\\[0mm]        

The following scenarios with two operators fail to accommodate the data with reasonable values of the WCs:\\[2mm]
\noindent $\blacktriangleright$  $\C_{qe} \,, \,\C_{\ell d} \rightarrow \,  \C_{9 \mu-e}^{\rm{NP}} = \C_{10 \mu-e}^{\rm{NP}}$   \,,   $\C_{9 \mu-e}^{\prime}  = -\C_{10 \mu-e}^{\prime}$\,: within this scenario it is not possible to accommodate both $R_{K^\ast}$ and $R_K$ simultaneously.\\[0.2cm]
\noindent $\blacktriangleright$  $\C_{\ell d} \,, \,\C_{ed} \rightarrow \,   \C_{9 \mu-e}^{\prime} \,, \,\C_{10 \mu-e}^{\prime}$\,: again, it is not possible to accommodate both $R_{K^\ast}$ and $R_K$ simultaneously.\\[0.2cm]
\noindent $\blacktriangleright$   $\C_{qe} \,, \,\C_{ed} \rightarrow \,  \C_{9 \mu-e}^{\rm{NP}} = \C_{10 \mu-e}^{\rm{NP}}$   \,,   $\C_{9 \mu-e}^{\prime}  = \C_{10 \mu-e}^{\prime}$\,: this scenario cannot generate the needed deviation on $R_K$.\\[-1mm]

In summary, the explanation of the $R_K$ and $R_{K^*}$ anomalies within the SMEFT at the level of dimension-six operators requires the presence of $\C_{\ell q}^{(1)}$ and/or $\C_{\ell q}^{(3)}$.

\vspace{5pt}

It is remarkable that, besides the hints of LFUV in $b \to s \ell^+ \ell^-$ transitions, a series of anomalies have also been observed in $b \to s \mu^+ \mu^-$~\cite{Hurth:2014vma,Altmannshofer:2014rta,Descotes-Genon:2015uva}.  A plausible scenario is that the NP enters mainly through muons, thus explaining the deviations from the SM in $b \to s \mu^+ \mu^-$ and the observation of LFUV when comparing muon and electron decay modes.  We will adopt this hypothesis in the following.  

We notice that all the viable explanations of the $R_K$ and $R_{K^*}$ anomalies considered in Section~\ref{sec:implications} provide a good fit of the $b \to s \mu^+ \mu^-$ data~\cite{Hurth:2014vma,Altmannshofer:2014rta,Descotes-Genon:2015uva}.  This observation is non-trivial given that a large fraction of the $b \to s \mu^+ \mu^-$  decay observables included in these global analyses probe different combinations of the WCs in general.    Note also that having only the operator $\C_{9 \mu}^{\rm{NP}}$ of the WET, which alone provides a very good fit of $b \to s \mu^+ \mu^-$ data, requires at least two SMEFT operators of the Warsaw basis, $\C_{\ell q}^{(1)}$ (or $\C_{\ell q}^{(3)}$) and $\C_{qe}$.  Other benchmark scenarios of the WET that provide a good fit, for instance $\C_{9 \mu}^{\rm{NP}}  = -\C_{9 \mu}^{\prime},\C_{10 \mu}^{\rm{NP}}  = \C_{10 \mu}^{\prime}$ are more involved to realize within the SMEFT due to the constraints imposed by electroweak gauge symmetry.

\vspace{5pt}

In Table~\ref{tab:WC_values} we use the result from the global fit to $b\to s\ell\ell$ in~\cite{Capdevila:2017bsm} to give the corresponding bounds on the WCs for the scenarios that can accommodate the $R_K$ and $R_{K^*}$ anomalies.  The involved WCs are $\mathcal{O}(1)$ for $\Lambda \sim 30$~TeV.
The result of the global fit in the scenario $(\C_{\ell q}^{(1,3)}, \C_{\ell d})$ is shown in Figure~\ref{fig:bounds} as red dashed contours.
\begin{center}
\renewcommand*{\arraystretch}{1.8}
\begin{table*}[ht]
\hfill{}
\begin{tabular}{c||c|c|c}
& \multicolumn{3}{c}{Fit from $b\to s\ell\ell$ observables} \TBstrut\\
\hline
Operator(s) $\times\, (30\, \mathrm{TeV}/\Lambda)^2$ & Best fit & $1\,\sigma$ & $2\,\sigma$ \TBstrut\\
\hline
\hline
$\C_{\ell q}^{(1,3)}$ & $0.95$ & $[0.75,1.14]$ & $[0.56,1.36]$ \TBstrut\\\rowcolor{CGray}
$(\C_{\ell q}^{(1,3)},\C_{qe})$ & $(1.03,0.80)$ & $\big([0.89,1.18],[0.61,0.98]\big)$ & $\big([0.74,1.32],[0.42,1.17]\big)$ \TBstrut\\
$(\C_{\ell q}^{(1,3)},\C_{\ell d})$ & $(1.02,-0.33)$ & $\big([0.80,1.23],[-0.54,-0.12]\big)$ & $\big([0.59,1.44],[-0.75,0.10]\big)$ \TBstrut\\\rowcolor{CGray}
$(\C_{\ell q}^{(1,3)},\C_{ed})$ & $(1.02,0.20)$ & $\big([0.81,1.22],[-0.00,0.41]\big)$ & $\big([0.60,1.43],[-0.21,0.62]\big)$ \Tstrut\\
\end{tabular}
\hfill{}
\caption{\small \sf  Constraints on the SMEFT WCs obtained from the global fit to $b\to s\ell\ell$ in terms of the WET operators from~\cite{Capdevila:2017bsm}.}\label{tab:WC_values}
\end{table*}
\end{center}

\section{Operator mixing effects}
\label{sec:constraints}

\subsection{Indirect contributions to $\mathbf{R_K}$ and $\mathbf{R_{K^\ast}}$}
\label{sec:myf}

The SMEFT WCs in the previous equations, given at $\mu = \mew$, can be
obtained in terms of their values at the NP scale $\Lambda$ by means
of the SMEFT RGEs~\cite{Jenkins:2013zja,Jenkins:2013wua,Alonso:2013hga}.
Using a first leading log approximation we find
\begin{widetext} 
\begin{align}  
[\C_{\ell q}^{(1)}(\mew)]_{aa23} =& \;  [\C_{\ell q}^{(1)}(\Lambda)]_{aa23}  -  \frac{y_t^2 \lambda_t^{sb}}{16 \pi^2}  \log \left( \frac{\Lambda}{\mew} \right) \left(      [\C_{\varphi \ell}^{(1)}(\Lambda)]_{aa}  -   [\C_{\ell u}(\Lambda)]_{aa33}     \right)\,,  \nonumber \\[0.2cm]
[\C_{\ell q}^{(3)}(\mew)]_{aa23} =& \;  [\C_{\ell q}^{(3)}(\Lambda)]_{aa23}  +  \frac{ y_t^2 \lambda_t^{sb}}{16 \pi^2}  \log \left( \frac{\Lambda}{\mew} \right) \left(     [\C_{\varphi \ell}^{(3)}(\Lambda)]_{aa}    \right) \,, \nonumber \\[0.2cm]
[\C_{qe}(\mew)]_{23aa} =& \;   [\C_{qe}(\Lambda)]_{23aa} -  \frac{y_t^2 \lambda_t^{sb} }{16 \pi^2}  \log \left( \frac{\Lambda}{\mew} \right) \left(     [\C_{\varphi e}(\Lambda)]_{aa}    -   [\C_{eu}(\Lambda)]_{aa33}   \right) \,,
\label{eqEWLambda} \\[0.2cm]
[\C_{\ell d}(\mew)]_{aa23} =&\;  [\C_{\ell d}(\Lambda)]_{aa23} \,, \nonumber \\[0.5cm]
[\C_{ed}(\mew)]_{aa23} = & \; [\C_{ed}(\Lambda)]_{aa23}  \,. \nonumber
\end{align}  \label{rete}
\end{widetext}
 We have made use of \textit{top
  dominance} assumptions, this is, we have only kept Yukawa terms
including $y_t= \sqrt{2} m_t/v\sim 1$, the top quark Yukawa coupling, neglecting
other Yukawa-driven terms.      We note that these expressions agree very
well with precise numerical calculations when the dominant terms are
the direct (tree-level) ones, while they may deviate slightly when the
one-loop induced terms dominate due to a non-negligible effect coming
from the running of the top Yukawa coupling. In the following we only
take them as guiding tool and obtain all our numerical results using
\dsix~\cite{Celis:2017hod}. We observe that, in principle, it is possible to achieve an explanation of the $R_{K,K^*}$ anomalies via operator mixing effects with a NP scale $\Lambda \sim 1$~TeV and WCs of $\mathcal{O}(1)$. Specifically, by generating $[\C_{\ell u}(\Lambda)]_{2233} \sim -1$, $[\C_{\varphi \ell}^{(1)}(\Lambda)]_{22}  \sim 1$, or $[\C_{\varphi \ell}^{(3)}(\Lambda)]_{22}  \sim - 1$.     However, we will see later that the possibility of $[\C_{\varphi \ell}^{(1,3)}(\Lambda)]_{22}$ is ruled out by experimental data.   For the interesting scenario, $[\C_{\ell u}(\Lambda)]_{2233}$, we include in Appendix~\ref{secClu} the one-loop matching corrections at the electroweak scale and compare these to the leading RGE contribution presented previously.

\subsection{Complementary constraints}
\label{sec:myf2}

Starting with the {\it exact} combination of Wilson coefficients at
the scale $\Lambda$ that generate the operators
$[Q_{\ell q}^{(1,3)}]_{2223}$ at the EW scale and nothing else, will be enough to
explain the anomalies as discussed in the previous section, with no
contributions to other very constrained observables. However, this
requires a fine tuning that will nevertheless be broken by loop
effects.

From the list of operators generated at the EW scale, some of them are
very strongly constrained because they contribute at tree level to
observables that are very precisely measured. The most relevant
observables in the case at hand are flavour-physics tests on
lepton-flavour universality, and EW precision
observables, see for instance~\cite{Boucenna:2016wpr,Boucenna:2016qad}. These
constraints will be the main potential obstacle to a coherent
explanation of the anomalies consistent with SM gauge invariance.

In this section we analyse the implications of the WCs
required to explain the anomalies in other low-energy observables. 
In particular we focus on the bounds from other LFUV observables 
and from Electroweak Precision Data (EWPD). We separate the
discussion in two cases: when the operators that explain the
anomalies are generated at tree-level and 
when they are induced at one-loop.

\begin{table}
\centering
\renewcommand*{\arraystretch}{1.5}
\begin{tabular}{@{}l||cc@{}}
WC ($\mu = \Lambda$) & $R_K$ and $R_{K^\ast}$ & Constraints \TBstrut\\[5pt]
\hline
\hline
\\[-5pt]
$\left[\C_{\ell q}^{(1)}\right]_{2223}$ & \cmark & No relevant constraints \TBstrut\\[5pt]
$\left[\C_{\ell q}^{(3)}\right]_{2223}$ & \cmark & No relevant constraints \TBstrut\\[5pt]
$\left[\C_{\varphi \ell}^{(1)}\right]_{22}$ & \xmark & Excluded due to EWPD \TBstrut\\[5pt]
$\left[\C_{\varphi \ell}^{(3)}\right]_{22}$ & \xmark & Excluded due to EWPD \TBstrut\\[5pt]
$\left[\C_{\ell u}^{\phantom{(1)}}\right]_{2233}$ & \cmark & No relevant constraints \TBstrut\\[5pt]
\end{tabular}
\caption{ \small \sf SMEFT operators at $\mu = \Lambda$ that can potentially explain the anomalies. The first two WCs contribute to $R_K$ and $R_{K^\ast}$ at tree-level while the last three contribute at the one-loop level. We find that $[\C_{\varphi \ell}^{(1,3)}]_{22}$ actually cannot work due to constraints from EWPD.}
\label{tab:summaryLam}
\end{table}

\bigskip
\noindent $\blacktriangleright$ \textbf{Tree-level generated operators:} First we focus on the 
observables that can give a direct constraint
on the operators given in Table~\ref{tab:WC_values}. As noted in 
Refs.~\cite{Buras:2014fpa,Calibbi:2015kma}, the operators $Q_{\ell q}^{(1,3)}$ could 
modify the ratio $R_{K^{(*)}}^{\nu\nu} = \Gamma(B \to K^{(*)}  \nu \bar \nu )/\Gamma(B \to K^{(*)}  \nu \bar \nu )_{\rm{SM}}$. Moreover, the WC  $\C_{\ell q}^{(3)}$ 
also affects the LFUV ratio $\Gamma_{B\to D^{(*)} \mu\nu}/\Gamma_{B \to D^{(*)} e\nu}$.
However we find that the contributions to these observables are always
below the experimental sensitivity. This result is consistent with
the analysis done in Ref.~\cite{Feruglio:2016gvd}. We do not find 
any other direct constraint on these scenarios. Furthermore,
we also consider the case where the relevant operators explaining
the anomalies are generated at the NP scale and use \dsix~\cite{Celis:2017hod} to obtain
the pattern of RGE-induced operators. We find that the new WCs 
generated in the running are sufficiently small to avoid the
experimental constraints from EWPD and LFUV observables.

\bigskip
\noindent $\blacktriangleright$ \textbf{One-loop induced operators:}
We now consider the possibility that the operators generated at the NP scale are
not among the ones that can explain the anomalies directly.
In this case the relevant contributions can still be generated through renormalization-group effects. Due to the loop suppression, the size of the WCs
necessary to account for the anomalies
should be larger and/or the NP scale should be lower, yielding
more interesting bounds at low energies. In fact, requiring WCs to be $\mathcal{O}(1)$ or smaller implies $\Lambda \lesssim \mathcal{O}(1)$~TeV in this case.  We find that among the
three possible scenarios, the ones based on $\C_{\varphi \ell}^{(1,3)}$
are excluded by EWPD since they induce excessively 
large modifications to the $W$ mass and/or the $Z$ couplings. In particular 
the required value of $\C_{\varphi \ell}^{(3)}$  is well beyond the allowed value from the bound on the
$W$ mass, while $\C_{\varphi \ell}^{(1)}$ induces a large contribution
to $Z \to \mu^+ \mu^-$ that is excluded by the LEP-I measurements, and to $\C_{\varphi D}$ (the WC of $Q_{\varphi D} = (\varphi^{\dag}   D^{\mu}   \varphi )^* (\varphi^{\dag}  D_{\mu}  \varphi )$) through the running which is also constrained
by the $W$ mass~\cite{Efrati:2015eaa,Falkowski:2015krw}. On the other hand we find that the
scenario where $\C_{\ell u}$ is obtained at the NP scale remains 
as a viable candidate, with:
\begin{equation}
[\C_{\ell u}(\Lambda)]_{2233} \sim -1 \quad {\rm and} \quad \Lambda \sim 1\ {\rm TeV}\ .
\end{equation}
RGE evolution down to the electroweak scale generates in this case contributions to $[Q_{\varphi \ell}^{(1)}]_{22}$ together with the four-lepton operators $[Q_{\ell \ell}]_{22aa} = (\bar \ell_2 \gamma_{\mu}  \ell_2 ) (\bar \ell_a \gamma^{\mu} \ell_a)$ and  $[Q_{\ell e}]_{22aa} = (\bar \ell_2 \gamma_{\mu}  \ell_2 ) (\bar e_a \gamma^{\mu} e_a)$, which are found to be well below the experimental limits~\cite{Efrati:2015eaa,Falkowski:2015krw}.   These findings are summarized in Table~\ref{tab:summaryLam}.

\section{Summary}
\label{sec:summary}

An increasing significance for new physics in $b\to s$ transitions is accumulating
since the first measurements of the $B\to K^\ast\mu\mu$ angular distribution 
by LHCb in~2013. While most of these observables are affected by hadronic uncertainties
(although meaningful theory predictions {\it can} and {\it have} been made), the smoking gun in this case is lepton-flavour universality violation (LFUV), hinted originally by the 2014 measurement
of $R_K$ in one large-recoil bin. A crucially important confirmation of such hints have appeared just recently with the LHCb measurement of $R_{K^\ast}$ in two large-recoil
bins. The importance of this measurement is that it is {\it complementary} to $R_K$
in regards to New Physics.

In this paper we have analysed the implications of these new measurements,
in terms of the SMEFT. Our conclusions on the required WCs at the scale
$\mu = \mew$ can be summarised as follows:\\[2mm]
\noindent $\blacktriangleright$ The $[ \C_{\ell q}^{(1,3)}]_{2223}$ coefficients play a crucial role in the explanation of the anomalies. All solutions (with one or two operators) require their presence to accommodate the LHCb measurements of $R_K$ and $R_{K^\ast}$.\\[2mm]
\noindent $\blacktriangleright$ The coefficients $\left[\C_{\ell d}\right]_{2223}$, $\left[\C_{qe}\right]_{2322}$ and $\left[\C_{ed}\right]_{2223}$ cannot explain the anomalies. The coefficient $\left[\C_{\ell d}\right]_{2223}$ fails since fixing it to get $R_K < 1$ one finds $R_{K^\ast} > 1$ in the central bin, contrary to what LHCb finds. The deviations induced by the other two coefficients are not large enough to match the measured values of $R_K$ and $R_{K^\ast}$.

\bigskip

Turning to our conclusions regarding the WCs at the UV scale, $\mu =
\Lambda$, they can be summarised as:\\[2mm]
\noindent $\blacktriangleright$ When the anomalies are explained with
operators that contribute to the $R_{K,K^\ast}$ ratios at tree-level
($[ \C_{\ell q}^{(1,3)}]_{2223}$), the resulting bounds are not
significant. In this case the NP scale can be as high as $\sim 30$\,-\,$50$ TeV
and still keep the WCs $\lesssim \mathcal{O}(1)$.\\[2mm]
\noindent $\blacktriangleright$  In contrast, when the anomalies are explained with operators that contribute via RGE operator-mixing effects ($[ \C_{\varphi \ell}^{(1,3)} ]_{22}$ and $[ \C_{\ell u} ]_{2233}$), the indirect bounds turn out to be very relevant. In fact, the coefficients $[ \C_{\varphi \ell}^{(1,3)} ]_{22}$ cannot explain the $R_{K,K^\ast}$ ratios since the required values are excluded by EWPD. For the $[ \C_{\ell u}]_{2233}$ coefficient no relevant constraints were found. In this case the NP scale must be very low once we assume $[ \C_{\ell u}]_{2233} \sim \mathcal{O}(1)$: $\Lambda \lesssim 1$ TeV, making this scenario potentially testable by other experimental means.

\vspace{10pt}

If confirmed, the violation of lepton flavour universality would have far-reaching consequences, implying the existence of new physics at energies relatively close to the TeV scale. In our analysis we have identified the crucial operators that a specific NP model would have to induce in order to be able to explain the $R_{K,K^\ast}$ anomalies. These minimal requirements can be regarded as a general guideline for model building.
In addition, when combining these measurements with all $b\to s\ell\ell$ data a consistent pattern arises (see Fig.~\ref{fig:bounds}), with the NP scenarios considered in this paper favoured with respect to the SM hypothesis by around 5 standard deviations, and with a high goodness of fit~\cite{Capdevila:2017bsm}.
We look forward for measurements of lepton-flavour universality-violating ratios at {\it low hadronic recoil}, as well as of other ratios such as $R_\phi$ and $R_{X_s}$, clean observables such as $Q_5$, and improved measurements with increased statistics.

\section*{Acknowledgements}
\linespread{1}\selectfont
{\small
 The work of A.C. is supported by the Alexander von Humboldt
Foundation.      The work of J.F. is supported in part by the Spanish
Government, by Generalitat Valenciana and by ERDF funds from the EU
Commission [grants FPA2011-23778,FPA2014-53631-C2-1-P,
  PROMETEOII/2013/007, SEV-2014-0398]. J.F. also acknowledges
VLC-CAMPUS for an ``Atracci\'o de Talent''
scholarship. A.V. acknowledges financial support from the ``Juan de la
Cierva'' program (27-13-463B-731) funded by the Spanish MINECO as well
as from the Spanish grants FPA2014-58183-P, Multidark CSD2009-00064,
SEV-2014-0398 and PROMETEOII/ 2014/084 (Generalitat Valenciana).
J.V. is funded by the Swiss National Science Foundation and
acknowledges support from Explora project FPA2014-61478-EXP.
}

\appendix

\section{Fermion mass basis}
\label{secfba}

After EW symmetry breaking, the Warsaw-basis operators are rotated to
the fermion mass basis by performing unitary transformations of the
fermion fields that diagonalise the fermion mass matrices,
\begin{align} \label{eq:rotations}
u_L \to& V_{u_L} u_L \,, \quad  d_L \to V_{d_L} d_L \,, \quad   u_R \to V_{u_R} u_R \,,  \nonumber \\
e_L \to& V_{e_L} e_L \,, \quad  e_R \to V_{e_R} e_R \,, \quad   d_R \to V_{d_R} d_R  \,.
\end{align}
In this way
\begin{align}
m_{\psi}^{\rm{diag}}  \equiv V_{\psi_L}^{\dag} \, m_{\psi}  \, V_{\psi_R} \,,
\end{align}
where
\begin{equation}
m_{\psi} = \frac{v}{\sqrt{2}} \left( \Gamma_\psi - \frac{1}{2} \frac{v^2}{\Lambda^2}\C_{\psi \varphi} \right)\,,
\end{equation}
is a diagonal and positive matrix corresponding to the physical
fermion masses. Here $\psi = \{ u,d,e \}$, $\Gamma_\psi$ is the Yukawa of the fermion $\psi$,
and $\C_{\psi\varphi}$ are the WCs of the SMEFT operators $Q_{\psi \varphi} = \left(
\varphi^\dagger \varphi \right) G_\psi$ that correct
the SM Yukawa operators $G_\psi$. We note
that these definitions imply that the CKM matrix is given by $V =
V_{u_L}^\dagger V_{d_L}$. This leads to the following
relations for the coefficients appearing in Eq.~\eqref{eqmatchgre}~\cite{Aebischer:2015fzz}:
\begin{align}\label{eq:tildeop}
\big[\tilde\C_{\ell q}^{(1)}\big]_{aa23} & = \big[ \C_{\ell q}^{(1)} \big]_{aamn} \left[ V_{d_L}^\dagger \right]_{2m} \left[ V_{d_L} \right]_{n3} \, , \nonumber \\
\big[\tilde\C_{\ell q}^{(3)}\big]_{aa23} & = \big[ \C_{\ell q}^{(3)} \big]_{aamn} \left[ V_{d_L}^\dagger \right]_{2m} \left[ V_{d_L} \right]_{n3} \, , \nonumber \\
\big[\tilde\C_{qe}\big]_{23aa} & = \big[ \C_{qe} \big]_{mnaa} \left[V_{d_L}^\dagger\right]_{2m} \left[V_{d_L}\right]_{n3} \, , \nonumber \\
\big[\tilde\C_{\ell d}\big]_{aa23} & = \big[\C_{\ell d}\big]_{aamn} \left[V_{d_R}^\dagger\right]_{2m} \left[V_{d_R}\right]_{n3} \, , \nonumber \\
\big[\tilde\C_{ed}\big]_{aa23} & = \big[\C_{ed}\big]_{aamn} \left[V_{d_R}^\dagger\right]_{2m} \left[V_{d_R}\right]_{n3} \,.
\end{align}
Throughout the paper we will assume that the WCs are given in the weak
basis where $V_{d_L}=\mathds{1}$ and $V_{d_R,u_R}=\mathds{1}$.

\section{One-loop contribution from $\C_{\ell u}$}
\label{secClu}

Assuming that the operator $[Q_{\ell u}]_{2233}$ is generated at the high scale $\Lambda$ by the new dynamics, we can obtain the complete one-loop result for the semileptonic WCs of the WET $\C_{9,10}^{\rm{NP}}$ using the results in~\cite{Aebischer:2015fzz} for the matching at the EW scale.    The final result, including the leading RGE contribution and the finite parts from the one-loop matching at the EW scale, reads  
\begin{align} \label{myspeq}
\C_{9 \mu}^{\rm{NP}}  \simeq  \frac{1}{s_W^2}  \frac{v^2}{\Lambda^2}  \frac{x_t}{8}    [\tilde \C_{\ell u}(\Lambda)]_{2233}   \left[    \log \left(  \frac{\Lambda}{M_W}\right)     + I_0(x_t)          \right] \,,
\end{align}
with $x_t = m_t^2/M_W^2$, $s_W = \sin \theta_W$ and the one-loop function $I_0(x_t) \simeq -0.71$ obtained from~\cite{Aebischer:2015fzz},
\begin{align}
I_0(x_t) = \frac{x_t - 7}{  4  (1- x_t)}   - \frac{   x_t^2 - 2 x_t + 4 }{  2(x_t -1)^2 }  \log(x_t)  \,.
\end{align}
In this scenario $\C_{10 \mu}^{\rm{NP}}  = -  \C_{9 \mu}^{\rm{NP}}$.    The relevant WC in the fermion mass basis is given by $[\tilde\C_{\ell u} ]_{2233}= [\C_{\ell u}]_{22mn} [V_{u_R}^\dagger]_{3m} [V_{u_R}]_{n3}$. We included in \eqref{myspeq} the finite parts that scale with the top-quark Yukawa coupling.    These cancel the $\mew$ scale dependence of the leading RGE contribution presented in Sec.~\ref{sec:myf}.    We find that for matching scales $\mew$ close to $m_t$, the NLO correction vanishes to a good approximation and the leading RGE contribution dominates, see~\cite{Bobeth:2017xry} for similar observations.  

\end{document}